\documentclass[twocolumn]{aastex631}

\usepackage{xspace}

\shortauthors{Jiang et al.}
\shorttitle{JWST-MUSE Lyman-alpha}
\newcommand\AAA{\mathrm{\AA}}
\newcommand\lya{\mathrm{Ly\alpha}}
\newcommand\gal{A2744-z6Lya\xspace}

\newcommand\ionf[2]{[#1$\;${\scshape{#2}}]}    % ion forbidden transitions, i.e., [O III] = \ionf{O}{iii}
\begin{document}

\title{The Ly$\alpha$ non-detection by JWST NIRSpec of a strong Ly$\alpha$ emitter at z\ =\ 5.66 confirmed by MUSE}

\author[0009-0006-1483-4323]{Haochen Jiang}
\affiliation{Deep Space Exploration Laboratory / Department of Astronomy, University of Science and Technology of China, Hefei 230026, China}
\affiliation{School of Astronomy and Space Science, University of Chinese Academy of Sciences (UCAS), Beijing 100049, China}

\author[0000-0002-9373-3865]{Xin Wang}
\affiliation{School of Astronomy and Space Science, University of Chinese Academy of Sciences (UCAS), Beijing 100049, China}
\affiliation{National Astronomical Observatories, Chinese Academy of Sciences, Beijing 100101, China}
\affiliation{Institute for Frontiers in Astronomy and Astrophysics, Beijing Normal University, Beijing 102206, China}

\author[0000-0003-0202-0534]{Cheng Cheng}
\affiliation{Chinese Academy of Sciences South America Center for Astronomy, National Astronomical Observatories, CAS, Beijing 100101, China}

\author[0000-0002-7660-2273]{Xu Kong}
\affiliation{Deep Space Exploration Laboratory / Department of Astronomy, University of Science and Technology of China, Hefei 230026, China}
\affiliation{School of Astronomy and Space Science, University of Science and Technology of China, Hefei 230026, China}

\author[0009-0006-1255-9567]{Qianqiao Zhou}
\affiliation{Deep Space Exploration Laboratory / Department of Astronomy, University of Science and Technology of China, Hefei 230026, China}
\affiliation{School of Astronomy and Space Science, University of Chinese Academy of Sciences (UCAS), Beijing 100049, China}

\author[0009-0006-0596-9445]{Xiao-Lei Meng}
\affiliation{National Astronomical Observatories, Chinese Academy of Sciences, Beijing 100101, China}

\author[0000-0002-1336-5100]{Xianlong He}
\affil{School of Astronomy and Space Science, University of Chinese Academy of Sciences (UCAS), Beijing 100049, China}
\affiliation{School of Physics and Technology, Wuhan University (WHU), Wuhan 430072, China}

\author[0000-0001-5860-3419]{Tucker Jones}
\affiliation{Department of Physics and Astronomy, University of California Davis, 1 Shields Avenue, Davis, CA 95616, USA}

\author[0000-0003-4109-304X]{Kristan Boyett}
\affiliation{School of Physics, University of Melbourne, Parkville 3010, VIC, Australia}
\affiliation{ARC Centre of Excellence for All Sky Astrophysics in 3 Dimensions (ASTRO 3D), Australia}

\correspondingauthor{Xin Wang, Cheng Cheng, Xu Kong}
\email{xwang@ucas.ac.cn , chengcheng@nao.cas.cn , xkong@ustc.edu.cn}

\begin{abstract}
\noindent
The detections of Lyman-$\alpha$ ($\rm Ly\alpha$) emission in galaxies with redshifts above 5 are of utmost importance for constraining the cosmic reionization timeline, yet such detections are usually based on slit spectroscopy. Here we investigate the significant bias induced by slit placement on the estimate of $\rm Ly\alpha$ escape fraction ($f_{\rm esc}^{\mathrm{Ly\alpha}}$), by presenting a galaxy (dubbed \gal) at $z=5.66$ where its deep JWST NIRSpec prism spectroscopy completely misses the strong $\rm Ly\alpha$ emission detected in the MUSE data.
\gal exhibits a pronounced UV continuum with an extremely steep spectral slope of $\beta=-2.574_{-0.008}^{+0.008}$, and it has a stellar mass of $\mathrm{\sim10^{8.82}~M_\odot}$, a star-formation rate of $\mathrm{\sim8.35~M_\odot yr^{-1}}$ and gas-phase metallicity of $\mathrm{12+log(O/H)\sim7.88}$. The observed flux and rest-frame equivalent width of its Ly$\alpha$ from MUSE spectroscopy are 
$1.2\times \rm 10^{-16}\,erg~s^{-1}cm^{-2}$ and 75\,$\AAA$, equivalent to $f_{\rm esc}^{\mathrm{Ly\alpha}}=78\pm4\,\%$. However, its Ly$\alpha$ non-detection from JWST NIRSpec gives a 5-$\sigma$ upper limit of $<13\,\%$, in stark contrast to that derived from MUSE. To explore the reasons for this bias, we perform spatially resolved stellar population analysis of \gal using the JWST NIRCam imaging data to construct 2-dimensional maps of SFR, dust extinction and neutral hydrogen column density. We find that the absence of Ly$\alpha$ in the slit regions probably stems from both the resonance scattering effect of neutral hydrogen and dust extinction. Through analyzing an extreme case in detail, this work highlights the important caveat of inferring $f_{\rm esc}^{\mathrm{Ly\alpha}}$ from slit spectroscopy, particularly when using the JWST multiplexed NIRSpec microshutter assembly.
\end{abstract}


\section{Introduction} \label{Sec.1}

Lyman-$\alpha$ emitters (LAEs) are remarkable tracers of the physical properties of the intergalactic medium (IGM) in the high-redshift universe \citep[see ][for a recent review]{Dijkstra2014}. These objects are characterized by a large rest-frame equivalent width of the $\lya$ emission line, typically $\mathrm{EW_0(\lya)\gtrsim 20~\AA}$ \citep{Ouchi2020}. \citet{Partridge1967} first assumed that strong Ly$\alpha$ emission line can be detected in young galaxies, and now we know LAEs are either young star-forming galaxies (SFGs) or QSOs. The ionization of hydrogen atoms by young massive stars leads to the emission of Ly$\alpha$ photons after the recombination process. In fact, due to factors such as dust extinction and gas scattering, only a subset of star-forming galaxies can be detected as LAEs \citep{Shapley2003}. Those SFGs with weak-to-no $\lya$ emission are mostly Lyman break galaxies \citep[LBGs,][]{Steidel2010}. The comparison of physical properties between LAEs and LBGs is discussed in detail by \cite{Ouchi2020}.

Studying LAEs is crucial for understanding galaxy formation and cosmic reionization \citep{Robertson2022}.  
%Confirming their redshifts through spectroscopy with modest exposure times allows us to explore the properties of galaxies in the high-redshift universe. 
The properties of Ly$\alpha$ emission are closely linked to Lyman continuum and the escape fraction of ionizing photons, providing insights into the contribution of star-forming galaxies to reionization \citep{Verhamme2015,Verhamme2017}. Given their history of intense star formation, LAEs act as ionization sources, creating ionized bubbles large enough for Ly$\alpha$ photons to pass through \citep{Madau1999}. As galaxies continue to ionize their neutral surroundings, Ly$\alpha$ photons can escape from the IGM. Consequently, the evolving properties of LAEs across cosmic epochs can reflect the evolution of the IGM \citep{Saxena2023a,Simmonds2023,Jones2023,Witstok2023,Izotov2019}
%According to the ionization model, the majority of ionizing photons must come from more fainter and lower-mass galaxies\citep{Robertson2015,Finkelstein2019}. Besides, some simulation works suggest that Ly$\alpha$ emission from faint satellite galaxies contribute to the origin of Ly$\alpha$ halos(LAHs) \citep{Shimizu2010,Lake2015,MasRibas2016,MasRibas2017,Mitchell2021,Byrohl2021}. The Cold Dark Matter (CDM) model of structure formation also predicts such faint sources around massive and brighter galaxies\citep{HerreroAlonso2023}.

A conventional method for finding LAEs involves narrow-band (NB) imaging. By comparing with broadband images, LAE candidates appear notably bright, and photometric redshifts are necessary for further certification. Numerous works have identified lots of LAEs using this approach \citep[e.g., ][]{Ouchi2008,Zheng2016,Zheng2017,Guo2020}. In recent years, the Multi-Unit Spectroscopic Explorer (MUSE) integral field spectrograph on the Very Large Telescope (VLT) has become a prominent tool for searching Ly$\alpha$ emission at various redshifts \citep[e.g.,][]{Wisotzki2018,Leclercq2020}. The $\lya$ emission line has an asymmetric line profile because the blue wing is absorbed more intensely. This distinctive feature enables clear identification, setting it apart from other spectral lines. Following the launch of the James Webb Space Telescope (JWST), researchers have seized opportunities to detect a number of fainter LAEs with lower masses than previously possible \citep{PrietoLyon2023,Roy2023}. Observations from JWST have already provided more constraints on the escape fraction of Ly$\alpha$ and LyC photons \citep{Matthee2023,Lin2023,Roy2023,Mascia2023}. 
%consequently aiding us in gaining deeper insights into the physical mechanisms conducive to the escape of ionizing photons during the cosmic reionization era \citep{Matthee2023,Lin2023,Roy2023,Mascia2023}.

Thanks to JWST's outstanding sensitivity and broad wavelength coverage in the near-infrared, we can conduct a comprehensive investigation into the UV continuum slopes, stellar populations, Ly$\alpha$ emissions, and other properties of the LAEs during the epoch of reionization, using both integrated and spatially resolved analyses \citep{Curti2022,Carnall2022,Topping2023,Iani2023,Roy2023,Nanayakkara2023,Shen2023,GimenezArteaga2023}. The synergy between the sensitivity of JWST instruments and the power of lensing magnification allows breakthroughs in Ly$\alpha$ detections at fainter UV magnitudes. In this paper, we present the JWST observations and results from both photometric and spectral analyses of a MUSE-selected strong Ly$\alpha$ emitting galaxy \gal (RA = 00:14:21.8; Dec = -30:23:44.0) at $z=5.66$ in the Abell 2744 (A2744) gravitational lensing field.
Using the deep prism spectroscopy from the NIRSpec multiplexed microshutter assembly (MSA) \citep{Jakobsen2022}, we confirm the spectroscopic redshift of \gal via the detection of a prominent Lyman break and a series of rest-frame optical strong nebular lines. 
Utilizing JWST/NIRCam \citep{Rieke2005,Rieke2023} and HST/WFC3+ACS imaging across twelve bands spanning the wavelength range of 0.8-5$\mathrm{\mu m}$, we perform spatially resolved spectral energy distribution (SED) fitting. 
Surprisingly, the pronounced Ly$\alpha$ emission seen in MUSE is not detected in the NIRSpec prism data, resulting in a significant bias in estimating the Ly$\alpha$ escape fraction. This underlines an important caveat concerning the potential interpretation of the cosmic reionization history derived from Ly$\alpha$ detections based on slit spectroscopy.
% Then we explain how the placement of spectroscopic slit affects the measurement of Ly$\alpha$, which as we believe is an important consideration in Ly$\alpha$ observations.

This Letter is structured as follows. We briefly describe the observations in Section \ref{Sec.2}. In Section \ref{Sec.3} we describe emission diagnostics from NIRSpec, spatially resolved properties of the galaxy and the analysis of Ly$\alpha$ emission. We present the results in Section \ref{Sec.4} and summarize the main conclusions in Section \ref{Sec.5}. We adopt a standard cosmology with $\mathrm{\Omega_{m}}$ = 0.3, $\Omega_{\Lambda}$ = 0.7, $\mathrm{H_0}$ = 70 $\mathrm{kms^{-1}Mpc^{-1}}$, and a Chabrier \citep{Chabrier2003} initial mass function (IMF).

\section{Observations} \label{Sec.2}

\begin{figure*}[htb]
    \centering
    \includegraphics[width=0.95\linewidth]{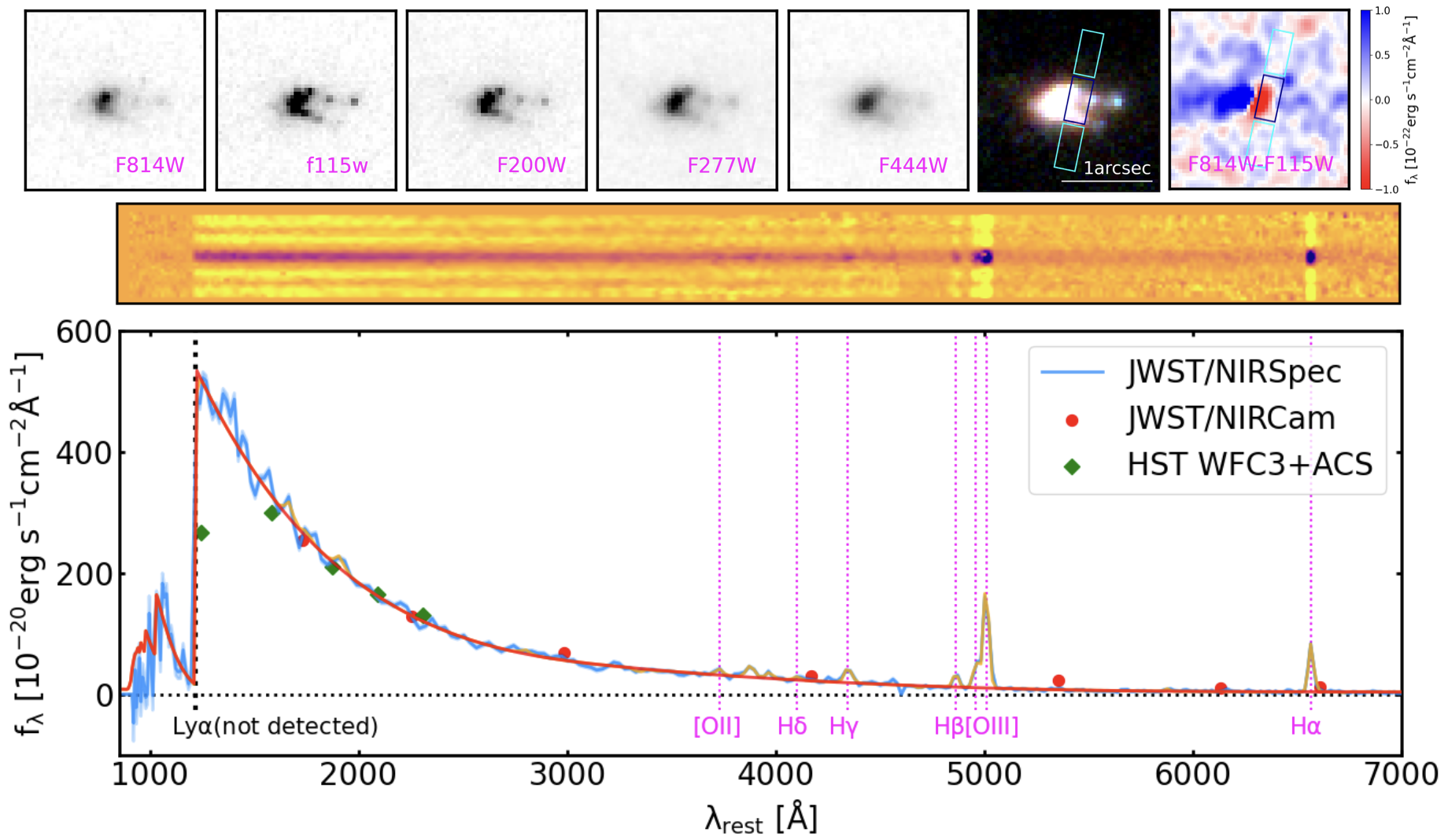}
    \caption{JWST+HST imaging and JWST/NIRSpec spectroscopy of \gal.
    {\bf Top}: \gal postage stamps in multiple filters and its color composite image produced from the NIRCam data. We used the F115W + F150W, F200W + F277W, and F356W + F444W images as blue, green, and red colors, respectively, to compose the color image. The far right image shows the color excess between F814W and F115W. The NIRSpec slits are overlaid on top of the color-composite image and the color-excess image, with the dark blue rectangle representing the central nodding position. {\bf Middle}: the unsmoothed 2D prism spectra combined from the three individual slit-dithering sequence, extracted using the up-to-date reduction software MSAEXP. We detect prominent emission features of the \ionf{O}{iii}$\lambda\lambda$4959,5007 doublets and a series of Balmer lines (i.e. H$\alpha$, H$\beta$, H$\gamma$), as well as the Lyman break at high significance.
    {\bf Bottom}: the optimally extracted 1D NIRSpec prism spectrum in blue with the 1-$\sigma$ error spectrum in the cyan-shaded band. The broad-band photometry measured from HST and JWST are represented by the green diamonds and the red circles, respectively.
    The red and orange lines correspond to our best-fit models for the source continuum and emission lines, respectively.}
    \label{Fig.1}
\end{figure*}

\begin{figure*}[htb]
    \centering
    \includegraphics[width=0.95\linewidth]{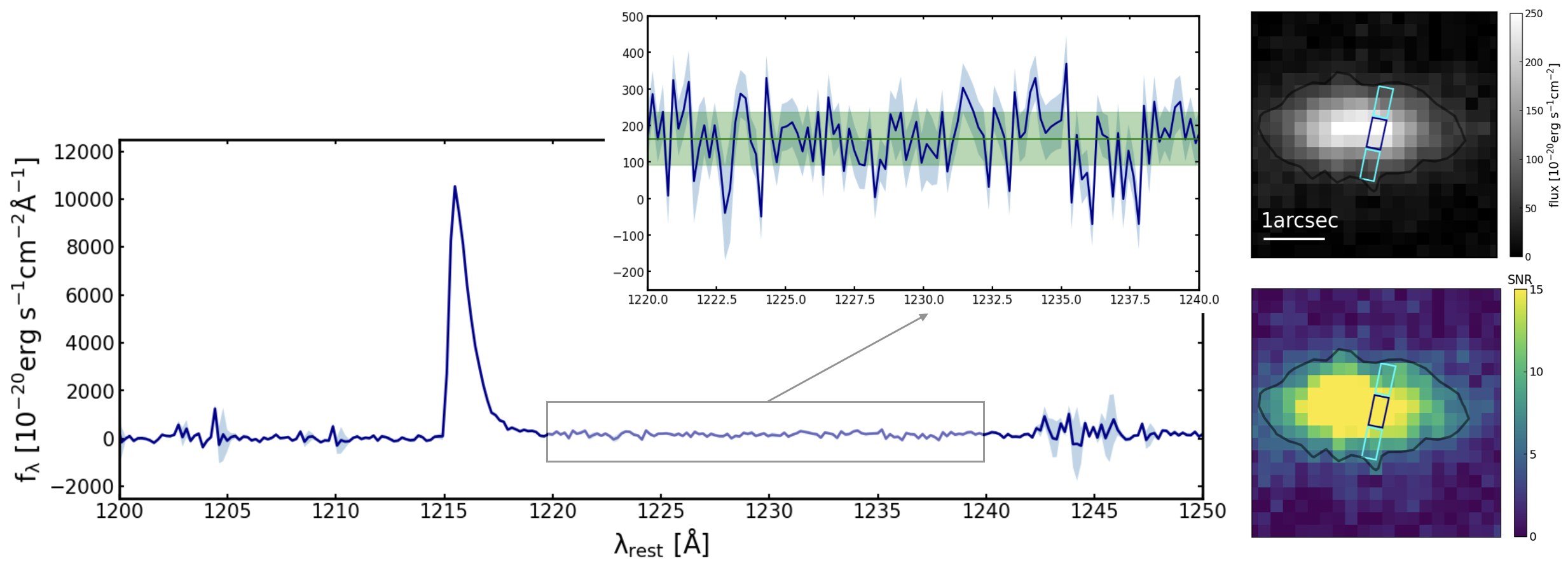}
    \caption{MUSE spectroscopy of \gal.
    {\bf Left}: The strong Ly$\alpha$ emission feature of \gal identified from the MUSE data. The 1-$\sigma$ error spectrum is represented as the cyan-shaded band. We zoom in on the continuum spectrum redward of Ly$\alpha$ in the inset, where the green line shows the average continuum flux, with the 1-$\sigma$ uncertainty in shadow. The continuum level measured from MUSE is consistent with that derived from JWST NIRSpec shown in Fig.~\ref{Fig.1}. 
    {\bf Right top}: pseudo-narrow band Ly$\alpha$ image, with the rest-frame wavelength in the range of  1214.9$\AAA$-1217.9$\AAA$. One pixel corresponds to 0.2 arcsec. {\bf Right bottom}: SNR map of the narrow band Ly$\alpha$ image. The spectroscopic slit in NIRSpec observations is also overlaid on top of both maps. The contour of SNR=5 are plotted in gray both on these two images.
    }
    \label{Fig.2}
\end{figure*}

\gal was reported as Ly$\alpha$ emitter from the VLT/MUSE observation in the A2744 field \citep[target ID: 8268, ][]{LaVieuville2020}. A2744 is one of the most famous galaxy cluster that has been covered by multi-wavelength. The wealth archive data especially the recent JWST image and spectroscopic observations, and the magnification by the galaxy cluster facilitate us to investigate the properties of \gal. We describe the available data used in this work below.

\subsection{JWST$\ \&\ $HST imaging}\label{2.1}

The JWST/NIRCam images of A2744 is reduced by GLASS project\footnote{\url{https://archive.stsci.edu/hlsp/glass-jwst}} \citep[PI: Tommaso Treu; PID: ERS-1324, GO-2561, and DD-2756,][]{Treu2022,Paris2023}. Details of the image data reduction can be found in \citep{Paris2023}. We make use of NIRCam images of \gal including F115W, F150W, F200W, F277W, F356W, F410M, F444W bands. The jwst pipeline version is 1.8.2 with the calibration file of {\sc jwst 1019.pmap}. All JWST images are psf-matched to F444W band on a common $0^{\prime\prime}.04~\mathrm{pixel}^{-1}$ scale. In this process, we also bin the images of F115W, F150W, and F200W two by two pixels for their original pixel scales are $0^{\prime\prime}.02~\mathrm{pixel}^{-1}$. 

For comprehensive photometry of this galaxy, we also include the publicly released HST imaging mosaics from the Hubble Space Telescope Frontier Fields (HFF) program \citep{Lotz2017} with ACS/WFC (F814W) and WFC3/IR (F105W, F125W, F140W, and F160W) bands. The photometric zero points and reference files of HST images are up to date.

In Fig.\ref{Fig.1}, we show images of six bands and a color composite image produced from these data. Besides, we use the stage I release of NIRCam catalogs in the Abell2744 central region. For our galaxy, we choose the catalog computed in 8$\times$ PSF apertures (diameters, which correspond to $1^{\prime\prime}.12$) and in the isophotal aperture. 

\subsection{JWST/NIRSpec spectrum}\label{2.2}

Our spectrum is acquired through NIRSpec multi-object spectroscopy (MOS) observations in the JWST Director Discretionary Time (DDT) program (PID 2756, PI: W. Chen; \cite{RobertsBorsani2022}), which obtained observations for targets located in the central regions of the Frontier Field galaxy cluster Abell 2744. DDT NIRSpec observations were carried out on October 23, 2022, using the CLEAR filter+PRISM configuration, which provides continuous wavelength coverage of 0.6-5.3 $\mu$m with R $\sim$ 30-300 spectral resolution. The on-source exposure time is 1.23 h. The original unsmoothed 2D prism spectrum is combined from the individual slit-dithering sequence, using the up-to-date reduction software MSAEXP. In this work, we focus on galaxy \gal. In Fig.\ref{Fig.1}, We show its 1D spectrum optimally extracted from the 2D spectral trace plotted in blue with the 1-$\sigma$ uncertainty in shadow, where prominent emission features of \ionf{O}{iii} and H$\alpha$ are observed.

\subsection{VLT/MUSE IFU data}\label{2.3}

The VLT/MUSE survey provides us with integral field unit (IFU) data. We downloaded the MUSE deep cube targeted at A2744 field at ESO website\footnote{\url{https://archive.eso.org/scienceportal/home}}. The cube we use is integrated by three programs \citep[PI: Richard, Johan;][]{LaVieuville2020}: 094.A-0115, 095.A-0181, 096.A-0496. The observed spectrum ranges from 475~nm to 935~nm, and the exposure time is 68400~s. Its FWHM effective spatial resolution is 0.64~arcsec, and 5-$\sigma$ limit magnitude limit is 25.96. In this cube, one pixel corresponds to 0.2~arcsec, and the interval of wavelength is 1.25~$\AAA$.
 
\section{Results and Analysis} \label{Sec.3}

\subsection{Integrated Analysis}\label{3.1}

We first employ the BAGPIPES \citep{Carnall2018,Carnall2019b} software to perform broadband SED fitting of our photometric data. Our basic assumptions include exponentially decaying star formation history (SFH) and the Calzetti \citep{Calzetti2000} dust extinction law with the visual extinction $\mathrm{A_V}$ in the range of 0-2. In this model, the decaying parameter $\tau$ ranges in 0.3-10 Gyr and stellar metallicity ($Z/Z_\odot$) ranges in 0-2.5. In addition, we conducted spectroscopic fitting using NIRSpec/PRISM spectrum. After the Chebyshev seventh order polynomial approximation, linear regression is used to measure the UV-continuum slope ($\beta$). We measure $\beta = -2.574_{-0.008}^{+0.008}$, which is consistent with the result of other work derived from photometry information \citep{Mascia2023}. The lens magnification is predicted from an online tool developed by \citet{Bergamini2023} \footnote{\url{http://bazinga.fe.infn.it:6005/SLOT}}, which is taken into account in the estimates of these physical properties when necessary. Derived by the extended model, the magnification factor ($\mu$) is $9.865_{-0.628}^{+0.850}$, and results on our source's stellar masses, SFR are reported in Table \ref{Table.1}. We find that \gal is a low mass ($\sim10^{8.8}~\mathrm{M_\odot}$), young stellar age ($\mathrm{\sim 50~Myr}$) SFG with SFR $\sim$ 8~$\mathrm{M_\odot yr^{-1}}$. 

\begin{table}[htb]
    \caption{Physical Properties of Galaxy \gal}
    \centering
    \small
    \renewcommand{\arraystretch}{1.2}
         
        \begin{tabular*}{\hsize}{@{}@{\extracolsep{\fill}}lc@{}}
            \hline
            \hline 
            Galaxy & \gal\\
            \hline
            R.A. (deg) & 3.590694 \\
            Decl. (deg) & -30.395542 \\
            $z_{\rm spec}$ & 5.66085 \\
            $\mu$ & $9.865_{-0.628}^{+0.850}$ \\
            $\beta$ & $-2.574_{-0.008}^{+0.008}$ \\
            \hline
            \multicolumn{2}{c}{Observed Emission-line Fluxes $\mathrm{[10^{-20}erg\,s^{-1}cm^{-2}]}$} \\
            $f$(\ionf{O}{iii}) & 7063.93$\pm$185.64\\
            $f(\mathrm{H\alpha})$ & 2429.84$\pm$108.83\\
            $f(\mathrm{H\beta})$ & 727.67$\pm$124.63\\
            $f$(\ionf{O}{ii}) & 387.42$\pm$183.27 \\
            \hline
            \multicolumn{2}{c}{Line Flux Ratios} \\
            O32 & $17.87\pm8.47$ \\
            R23 & $10.23\pm1.79$ \\
            H$\alpha$/H$\beta$ & $3.34\pm0.59$ \\
            \hline 
            \multicolumn{2}{c}{Rest-frame Equivalent Widths [$\AAA$]} \\
            $\mathrm{EW}$(\ionf{O}{iii}) & 1252.47$\pm$32.91\\
            $\mathrm{EW}(\mathrm{H\alpha})$ & 642.81$\pm$28.79\\
            $\mathrm{EW}(\mathrm{H\beta})$ & 117.94$\pm$20.20\\
            $\mathrm{EW}($\ionf{O}{ii}) & 21.15$\pm$10.00\\
            \hline
            \multicolumn{2}{c}{Nebular Emission Diagnostics} \\
            12+log(O/H) & $7.8778_{-0.0525}^{+0.0721}$\\
            $\mathrm{SFR^N\,[M_{\odot} \rm yr^{-1}]}$ & $8.231_{-2.135}^{+3.533}$\\
            $\mathrm{A_V^N}$ & $0.5626_{-0.3003}^{+0.3572}$\\
            \hline
            \multicolumn{2}{c}{Broadband Photometry SED fitting} \\
            log($M_*/M_{\odot}$) & $8.821_{-0.004}^{+0.009}$ \\
            $\mathrm{SFR^S\,[M_{\odot} \rm yr^{-1}]}$ & $8.352_{-0.100}^{+0.196}$ \\
            $\mathrm{sSFR^S}\,[10^{-8} \rm yr^{-1}]$ & $1.261_{-0.019}^{+0.040}$ \\
            Age\,[Myr] & $50.53_{-0.02}^{+0.05}$ \\
            \hline
            \multicolumn{2}{c}{Ly$\alpha$ Diagnostics} \\
            $f(\lya)\,\mathrm{[10^{-20}erg\,s^{-1}cm^{-2}]}$ & 12214$\pm$99 \\
            $\mathrm{EW_0(\lya)}\,[\AAA]$ & 75$\pm$33 \\
            $\mathrm{f_{esc}^{\lya}(MUSE)}$ & 0.783$\pm$0.035\\
            $5\sigma$ upper limit of $\mathrm{f_{esc}^{\lya}(NIRSpec)}$ & 0.134\\
            \hline
        \end{tabular*}
        
    \raggedright
    \textbf{Note.} The observed emission line fluxes are values before correction for dust extinction and lensing magnification. The values of $\mathrm{M_*}$ and SFR have been corrected for lens magnification. All uncertainties presented in this table correspond to 1-$\sigma$ confidence intervals.
    \label{Table.1}
\end{table}

\subsection{Rest-frame Optical Line Measurements}\label{sect:line_flux}

The spectrum extracted from NIRSpec shows a few measurable rest-frame optical emission lines, although it has a relatively low resolution for split by prism. To compensate for the slit loss, we adopt two different methods. On the one hand, we compare the best-SED-fitting results of spectral fitting with that of photometric fitting. Multiply the flux density of spectral fitting results by a scale factor which is 5 in our situation, these two results will approximately overlap at rest-frame ultraviolet band. On the other hand, in F444W image, the proportion of flux in the slit is 20\,\%. The result is consistent with the scale factor, demonstrating the robustness of our methods. Thus we are able to make corrections in subsequent analysis using the scale factor. 

We use the MSAEXP \footnote{\url{https://github.com/gbrammer/msaexp}} Python package to fit the spectrum and then obtain the continuum line (shown in Fig.\ref{Fig.1}, the red line), and then subtract it from the total spectrum as the flux of emission lines (the orange line in Fig.\ref{Fig.1}). We choose LMFIT \footnote{\url{https://lmfit.github.io/lmfit-py/}} package to calculate fluxes and equivalent widths of these emission lines, including \ionf{O}{ii}, \ionf{O}{iii}$\lambda\lambda$4959,5007 doublets, H$\alpha$ and H$\beta$. Following previous works \citep[][]{Wang2017,Wang2019,Wang2020,Wang2022,Wang2022a}, we employ the Bayesian method to jointly constrain the ionized gas metallicity $(12+\log(\mathrm{O/H}))$, nebular dust extinction $\mathrm{(A_V^N)}$, and de-reddened $\mathrm{H}\beta$ line flux $(f_{\mathrm{H}\beta})$, based on the strong line calibration \citep{Bian2018}. Besides, the instantaneous star formation rate $(\mathrm{SFR^N})$ is estimated from $f_{\mathrm{H}\beta}$ assuming the \citet{Kennicutt1998} calibration and the Balmer decrement ratio of $\mathrm{H}\alpha/\mathrm{H}\beta=2.86$ scaled to the \citet{Chabrier2003} IMF, i.e., $\mathrm{SFR^N} = 4.65\times10^{-42} \frac{\mathrm{L(H\beta)}}{\mathrm{[erg\,s^{-1}]}}\times 2.86\,[\mathrm{M_\odot yr^{-1}}]$. These results are all listed in Table \ref{Table.1}. 
%The result of $\mathrm{SFR^N}\sim8.35\mathrm{M_\odot yr^{-1}}$ corresponds to that from photometry $\mathrm{\sim8.23M_\odot yr^{-1}}$, confirming strong star-forming activity in \gal.

\begin{figure*}[htb]
    \centering
    \includegraphics[width=0.9\linewidth]{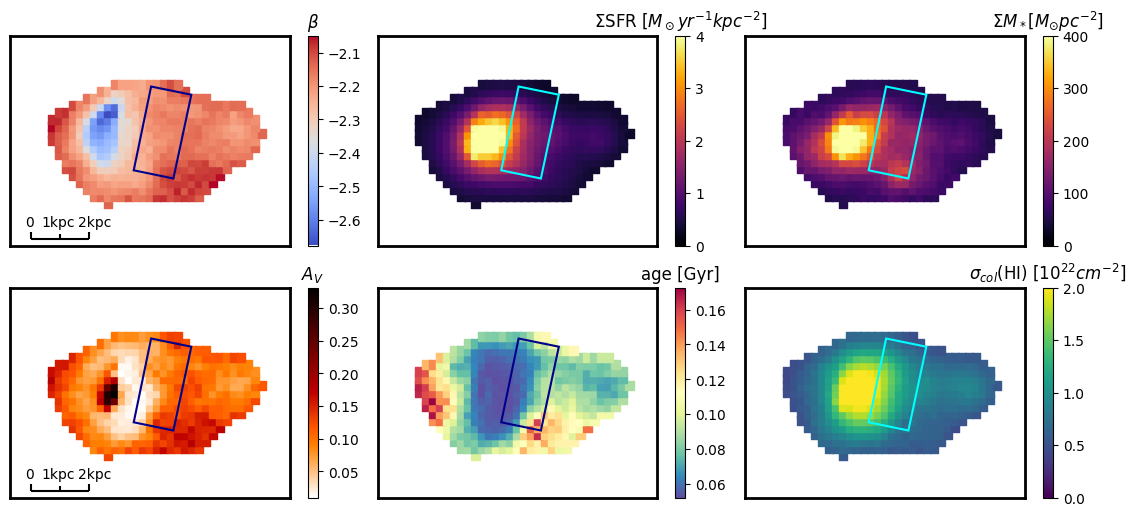}
    \caption{Spatially resolved physical properties of \gal at $z = 5.66$ derived using our 2D SED modeling.
    The panels in the top row, from left to right, show the 2D maps of UV slope, SFR surface density, and stellar mass surface density.
    The panels in the bottom row show the 2D maps of visual dust extinction $\mathrm{A_V}$, the mass-weighted stellar age, and the gas column density, based on the extended Schmidt law \citep{Shi2018} and the result of \cite{Markov2022}. One pixel corresponds to 0.04 arcsec. The NIRSpec slit in the central nodding position is marked in either dark blue or cyan in each panel.}
    \label{Fig.3}
\end{figure*}

\begin{figure}[htb]
    \centering
    \includegraphics[width=0.9\linewidth]{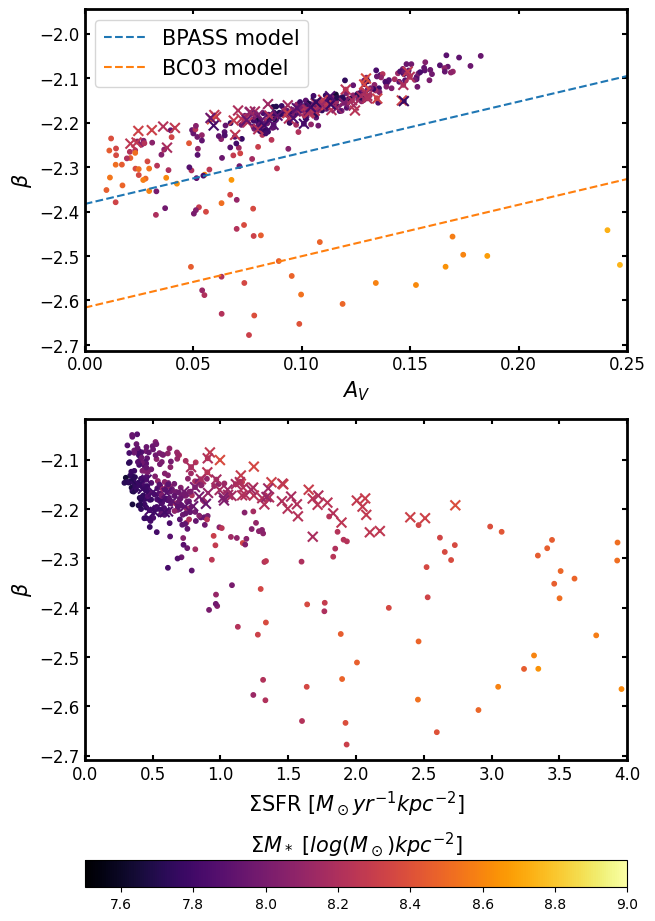}
    \caption{Correlations among $\beta$, $\mathrm{A_V}$ and $\Sigma$SFR derived from our spatially resolved analysis. We highlight the pixels within the central slit using crosses. The symbols are color-coded in $\Sigma M_*$.
    The dashed lines in the top panel represent the relations between $\beta$ and $\mathrm{A_V}$ obtained by the HDUV survey \citep[as summarized in Table~3 of][]{Reddy2018}, assuming the 0.14 $Z_{\odot}$ BPASS and 1.4 $Z_{\odot}$ BC03 stellar population models. The regions of \gal covered by the slit follow these general trends and are more compatible with the $\beta$-$\mathrm{A_V}$ relation given by the low-metallicity BPASS model, consistent with its metallicity measurement from NIRSpec data (see Table~\ref{Table.1}).}
    \label{Fig.4}
\end{figure} 

\subsection{Ly$\alpha$ Emission}\label{3.3}

Narrowband imaging is a classical technique for detecting strong Ly$\alpha$ emission. Using the MUSE data cube, we construct a pseudo-narrowband Ly$\alpha$ image (Fig.\ref{Fig.2}) of \gal. This involves centering on the galaxy's position and stacking images at wavelengths corresponding to the Ly$\alpha$ emission line, with a spectral bandwidth ranging from rest-frame 1214.9\,$\AAA$ to 1217.9\,$\AAA$ for our source. We show it in Fig.\ref{Fig.2}, together with the signal-to-noise ratio (SNR) of the narrowband image. To calculate the total $\lya$ flux, we select pixels with SNR\,$>5$ (shown as pixels within the contour in Fig.\ref{Fig.2}). Summing these pixels provides a spectrum containing the Ly$\alpha$ emission line and the adjacent continuum (Fig.\ref{Fig.2}). The average flux density within the wavelength range of 1220\,$\AAA$-1240\,$\AAA$ is computed as the flux density of the continuum. Subsequently, we determine the flux and equivalent width (EW) of Ly$\alpha$, as shown in Table \ref{Table.1}. The large EW suggests that \gal is a LAE at high redshift. 

The Ly$\alpha$ escape fraction is defined as $f_{\mathrm{esc}}^\lya=\mathrm{L(Ly\alpha)/L_{int}(Ly\alpha)}$. Under case B recombination and taking 8.7 \citep{Henry2015} as the ratio of $\mathrm{L_{int}(Ly\alpha)}$ to $\mathrm{L_{int}(H\alpha)}$, we calculate the total intrinsic Ly$\alpha$ flux. Therefore, we obtain a precise escape fraction of Ly$\alpha$ photons, which is 0.783$\pm$0.035 from MUSE observation. 

With the blue UV continuum of this galaxy, we observe a spatial offset between $\lya$ emission and the UV continuum. This offset is also noted in \citet{Lemaux2021}. Their work points out that the offset of galaxies at z$\sim$6 is quite common and it is concluded that the offset between $\lya$ emission and UV continuum is $\sim$0.6 kpc at z=6, equivalent to half the width of a 0.2\,arcsec wide slit in NIRSpec MOS observation. Fig.\ref{Fig.1} displays the image of color excess between F814W and F115W, illustrating this offset. For \gal, F814W well includes $\lya$ emission, and F115W captures the redder continuum. The conspicuous red area in slit region indicates a strong rest-frame UV continuum, while $\lya$ photons escape from the intense blue area near the slit. This spatial offset, consistent with the non-detection by NIRSpec, spans nearly one width of slit. Such offsets can introduce significant bias on $\lya$ escape fraction estimates, as we will discuss later in the case of \gal.

\subsection{Spatially Resolved Properties}

We use imaging data mentioned in \ref{2.1} and utilize the segmentation map of this source to mask background pixels and also remove pixels with SNR\,$<3$ in the F200W band. Then we conduct pixel-by-pixel SED fitting, fitting parameter settings the same as \ref{3.1} mentioned. Thus, we obtain fitted results including stellar mass ($\mathrm{M_*}$), star formation rate and dust extinction $\mathrm{A_V}$. In the posterior, we take the 50th percentile values as the result of the corresponding physical parameters to present. According to the relative position among the pixels, we plot two-dimensional spatially resolved maps, as shown in Fig.\ref{Fig.3}. These images reflect the distribution or relative strength of the physical parameters of our source. The surface density of gas mass is obtained from the extended Schmidt law \citep{Shi2018}, i.e. $\Sigma \mathrm{SFR = 10^{-4.76}\kappa_s\,(\Sigma M_*^{0.5}\Sigma M_{gas})^{1.09}}$, where $\Sigma \mathrm{SFR}$ is in $\mathrm{M_\odot yr^{-1}kpc^{-2}}$, $\mathrm{\Sigma M_*}$ and $\mathrm{\Sigma M_{gas}}$ are in $\mathrm{M_\odot pc^{-2}}$. Here we add a variable parameter $\kappa_s$, because starburst galaxies at high redshift locate above the KS law \citep{Heiderman2010,Ferrara2019}, and we adopt the result in \cite{Markov2022}, which is $\mathrm{log\,(\kappa_s)=1.43}$ for a galaxy at z\,=\,5.66. Thus we convert it to HI column density assuming the neutral hydrogen atoms dominate the gas budget of our galaxy. Based on the spatially resolved results, we calculate that the slit area ratios of SFR, stellar mass, and gas mass are all approximately 20\,\%. 

Furthermore, we draw scatter plots to show the distinct relationship between $\beta$, $\mathrm{A_V}$, and SFR directly. In Fig.\ref{Fig.4}, every dot represents every pixel in our maps, and the pixels in the region covered by the central NIRSpec slit are marked in ``$\times$''. Through the maps and scatter plots of $\beta$ and extinction $\mathrm{A_V}$, there is a strong correlation between these two physical parameters. \cite{Reddy2018} gave the relations of $\beta$ and $\mathrm{E(B-V)}$ by HDUV survey based on the Calzetti dust curve and the 0.14 $Z_{\odot}$  BPASS and 1.4 $Z_{\odot}$ BC03 stellar population models assuming different metallicities. We convert them into the distribution between $\beta$ and $\mathrm{A_V}$. We can see that the slope of this relation fits our observations well. The shift of the interception is primarily due to the different metallicities adopted in the stellar population models.

\section{Discussion}\label{Sec.4}

\subsection{Correction of Slit Loss}\label{4.1}

To study the impact of the placement of the spectroscopic slit, we show the position of the slit in the RGB image (Fig.\ref{Fig.1}). We can see that the slit region covers about one-third of the luminous area of the whole galaxy. 
Astrometry of MUSE datacube from default pipeline is offest from NIRCam images. To unify the data astrometry, we detect fifteen sources around \gal in a $40^{\prime\prime}\times40^{\prime\prime}$ area as the reference targets, and find the offsets of 0.4\,arcsec southward and 0.2\,arcsec eastward for the astrometry of MUSE relative to that of JWST/NIRCam.
The slits after shift are plotted in Fig.\ref{Fig.2}. In our calculation, Ly$\alpha$ flux where slit covers accounts for approximately 4\,\% of the total flux. 

In \ref{3.3} we have derived the escape fraction of Ly$\alpha$ obtained from MUSE data. Now we will give an upper limit for escape fraction using NIRSpec data. We calculate the average error of flux density in the rest-frame wavelength band of $1000-2000\,\AAA$. Then we take the five times the error as the upper limit of Ly$\alpha$ emission line flux. Combined with total intrinsic Ly$\alpha$ flux, we further derive the 5-$\sigma$ upper limit of Ly$\alpha$ escape fraction to be 0.134. We find that there exists a considerable gap compared to the result from MUSE. In the following content, we will discuss several factors that have contributed to this large bias.

\subsection{Resolution of PRISM Spectrum}\label{4.2}

The fact that our spectrum is split by the prism in NIRSpec with a relatively low resolution $R \sim100$, may also contribute to the non-detection of $\lya$ in our spectrum. For \gal, the rest-frame equivalent width EW$_0$ of $\lya$ is $\mathrm{\sim75\,\AAA}$. We compare it with other studies. \citet{Saxena2023a} reported a galaxy at z = 7.3 with an extremely large $\mathrm{EW_0}$ $\lya$ emission measuring 388\,$\AAA$, detected in both PRISM and grism spectra. In a study by \citet{Tang2023}, which included 21 galaxies at z\,$\gtrsim$\,7, two of them showed Ly$\alpha$ detection in PRISM with $\mathrm{EW_0}$ values of $\sim41.9\,\AAA$ and $\sim77.6\,\AAA$. Additionally, \citet{Saxena2023b} presented 16 LAEs at z\,$\sim5.8-8.0$ with $\mathrm{EW_0}(\lya)$ ranging in $25-350\,\AAA$, all detected in both PRISM and G140M spectra. 
While these studies have detected Ly$\alpha$ with lower equivalent widths than that of \gal, a direct comparison is challenging because our result does not account for slit loss. Assuming the previous correction for slit loss, the probable value is $\sim15\,\AAA$. Therefore, we cannot exclude the possibility that the low equivalent width of $\lya$ is not be distinguished by the PRISM resolution of 14\,$\AAA$ at $\lya$ wavelength.

\subsection{Star Forming Regions}\label{4.3}

As we can see in the RGB image in Fig.\ref{Fig.1}, the slit is located near the center of \gal. However, the region it covers is offset from the peaks of both the SFR and $\mathrm{M_*}$ surface density maps, which are to the left (i.e. east) of the slit. While the effect of off-center slit placement has been accounted for in the slit loss correction described in Sect.~\ref{sect:line_flux}, there are additional physical reasons that can result in the obstruction of Ly$\alpha$ escape on sub-galactic scales.
Firstly, since Ly$\alpha$ is a resonance line, the column density of HI gas toward a Ly$\alpha$ source (e.g. star-forming regions) is a key physical quantity that scatters the Ly$\alpha$ emission off the line of sight.
Secondly, dust can effectively attenuate the Ly$\alpha$ photons, if there exist sufficient dust grains around the star-forming regions.
According to our spatially resolved analysis shown in Fig.~\ref{Fig.3}, for the non-detection of Ly$\alpha$ emission from the NIRSpec slit of \gal, HI gas absorption and dust extinction both play an important role. 

What may be confusing is the strong Ly$\alpha$ flux in the MUSE image, especially in the regions covered by the NIRSpec slit. Two different instruments targeting the same area present completely different results. The key explanation to reconcile this is the spatial resolution of MUSE. For our data, it is $0^{\prime\prime}.64$, which is equal to three pixels in the image. We could totally imagine that the area with strong Ly$\alpha$ signal is not as large as shown in the NB image. Instead, it might be as large as the star-forming region in the spatially resolved maps. According to the SNR image (in Fig.\ref{Fig.2}), this interpretation can be well confirmed. Regions with high SNRs almost exactly coincide with star-forming regions. Therefore, We believe that the extended distribution of Ly$\alpha$ seen from MUSE is an artifact of the seeing-limited ground-based data. 
%A variety of works have demonstrated the point that $\lya$ halos of high redshift star-forming galaxies are extended \citep{Wisotzki2016,Leclercq2017,Erb2018}. 
The JWST NIRSpec IFU spectroscopy is required to accurately characterize the spatial extent of the Ly$\alpha$ emission from \gal.

\section{Summary and Conclusions}\label{Sec.5}

In this Letter, we present a $z=5.66$ LAE confirmed by MUSE IFU data with the absence of Ly$\alpha$ emission line in JWST NIRSpec prism spectrum, reiterating the strong bias caused by slit loss. 

With JWST NIRCam data, we conduct broadband photometry SED fitting, obtaining the stellar mass of $\sim\mathrm{10^{8.82}\,M_\odot}$, SFR of $\sim\mathrm{8.35\,M_\odot yr^{-1}}$ and further sSFR of $\sim\mathrm{1.26\times10^{-8}\,yr^{-1}}$. Besides, we use spectroscopic fitting and make emission diagnostics, giving a blue UV slope of $\beta\sim-2.574$, low metallicity of $\mathrm{12+log(O/H)\sim7.88}$ and instantaneous SFR of $\sim\mathrm{8.23M_\odot yr^{-1}}$. These observations are consistent with typical properties of LAEs, including low stellar mass and high sSFR. Based on the MUSE data cube, we construct the Ly$\alpha$ pseudo-narrowband image and spectrum containing strong $\lya$ emission feature of this galaxy. The observed flux of $\lya$ is $\mathrm{\sim1.22\times10^{-16}\,erg\ s^{-1}cm^{-2}}$ and $\mathrm{EW_0(Ly\alpha)=75\pm33\,\AA}$. Furthermore, we calculate Ly$\alpha$ escape fraction of $0.783\pm0.035$, and estimate the 5$\sigma$ upper limit of Ly$\alpha$ escape fraction of 0.134 using NIRSpec observation. This significant disagreement between the two $f_{\mathrm{esc}}^\lya$ is notable and worthy of caution.

We discuss the reasons for the chasm of Ly$\alpha$ escape fraction estimates given by different instruments in detail. To investigate the $\lya$ non-detection in NIRSpec, we present spatially resolved stellar population properties of \gal. It has a central region where SFR is significantly high, but the spectroscopic slit dislocated from it. What is more, both the strong gas scattering and dust extinction are likely to contribute to the prevention of the escape of Ly$\alpha$ photons. At the same time, we can not exclude the possibility that the low resolution of prism results in this fact, but even in this situation, the escaped Ly$\alpha$ flux may also be very weak. 

By analyzing an extreme case, this work points out the crucial caveat of deriving $f_{\mathrm{esc}}^\lya$ from slit spectroscopy. This strong bias results in a significant reduction of $f_{\mathrm{esc}}^\lya$ from $78\pm4~\%$ to a stringent upper limit of $<13~\%$ at 5-$\sigma$ confidence level. It is thus essential to take into account this bias when interpreting the reported Ly$\alpha$ escape fractions derived from slit spectroscopy, in particular with JWST NIRSpec MSA.

\begin{acknowledgments}
This work is based on observations made with the NASA/ESA/CSA James Webb Space Telescope, associated with the program JWST-DD-2756 (PI: Chen) and JWST-GO-2561 (PI: Labbe). 
The data were obtained from the Mikulski Archive for Space Telescopes at the Space Telescope Science Institute, which is operated by the Association of Uni- versities for Research in Astronomy, Inc., under NASA contract NAS 5-03127 for JWST.
We thank Charlotte Mason for useful discussion.
Xin Wang is supported by the Fundamental Research Funds for the Central Universities, and the CAS Project for Young Scientists in Basic Research, Grant No. YSBR-062. 
H.J. and X.K. are supported by the Strategic Priority Research Program of Chinese Academy of Sciences (Grant No. XDB 41000000), the National Science Foundation of China (NSFC, Grant Nos. 12233008, 11973038), the China Manned Space Project (No. CMS-CSST-2021-A07), the Cyrus Chun Ying Tang Foundations, the Frontier Scientific Research Program of Deep Space Exploration Laboratory, and the 111 Project for ``Observational and Theoretical Research on Dark Matter and Dark Energy" (B23042).
\end{acknowledgments}

%\appendix

\bibliography{JWST-MUSE}{}
\bibliographystyle{aasjournal}

\end{document}